\documentclass[twocolumn,floats,showpacs,superscriptaddress]{revtex4}
\usepackage{amssymb}
\usepackage{graphicx}
\usepackage{dcolumn}
\usepackage{amsmath}
\usepackage{bm}
\usepackage{epsfig}

\begin{document}

\title{Correlations in interacting systems with a network topology}
\author{S. N. Dorogovtsev}
\email{sdorogov@fis.ua.pt}
\affiliation{Departamento de F{\'\i}sica da Universidade de Aveiro, 3810-193 Aveiro,
Portugal}
\affiliation{A. F. Ioffe Physico-Technical Institute, 194021 St. Petersburg, Russia}
\author{A. V. Goltsev}
\email{goltsev@fis.ua.pt}
\affiliation{Departamento de F{\'\i}sica da Universidade de Aveiro, 3810-193 Aveiro,
Portugal}
\affiliation{A. F. Ioffe Physico-Technical Institute, 194021 St. Petersburg, Russia}
\author{J. F. F. Mendes}
\email{jfmendes@fis.ua.pt}
\affiliation{Departamento de F{\'\i}sica da Universidade de Aveiro, 3810-193 Aveiro,
Portugal}
\date{}

\begin{abstract}
We study pair correlations in cooperative systems placed on complex
networks. We show that usually in these systems, the correlations between
two interacting objects (e.g., spins), separated by a distance $\ell $,
decay, on average, faster than $1/(\ell z_{\ell })$. Here $z_{\ell }$ is the
mean number of the $\ell $-th nearest neighbors of a vertex in a network.
This behavior, in particular, leads to a dramatic weakening of correlations
between second and more distant neighbors on networks with fat-tailed degree
distributions, which have a divergent number $z_{2}$ in the infinite network
limit. In this case, only the pair correlations between the nearest
neighbors are observable. We obtain the pair correlation function of the
Ising model on a complex network and also derive our results in the
framework of a phenomenological approach.
\end{abstract}

\maketitle



\address{de Aveiro, \\
Campus Universit\'{a}rio de Santiago, 3810-193 Aveiro, Portugal\\
$^{2}$ A.F. Ioffe Physico-Technical Institute, 194021 St. Petersburg,
Russia}

\address{$^{1}$ Departamento de F\'{\i}sica, Universidade \\
de Aveiro, Campus Universit\'{a}rio de Santiago, 3810-193 Aveiro, Portugal\\
$^{2}$ Ioffe Physico-Technical Institute, 194021 St. Petersburg, Russia}


\section{\label{In}Introduction.}

The generic features of real-world networks (the Internet, the WWW,
biological, social and economical networks and many others) are a complex
organization of their connections and the small-world phenomenon \cite%
{s01,ab01a,dm01c,n03,w99,dm03,pvbook04,ajb99}. In the networks with the
small-world effect, a mean intervertex distance grows with a total number of
vertices, $N$, slower than any power of $N$, e.g., grows as $\ln N$.
Concerning economical and social networks, the small-world property is often
considered as an evidence for growing close interrelations and globalization.

Many of the real-world networks are formed by interacting objects and
demonstrate complicated dynamics. The dependence of the correlations between
interacting objects in systems with a network topology on time and distance
provides a useful information about the network dynamics. In the present
paper we discuss general properties of the correlations between a pair of
interacting objects on a complex network. Recall that in interacting systems
on lattices, with a few exceptions, pair correlations decrease exponentially
with distance apart from a critical point, where the decrease is power-law.
From a naive point of view, one might expect that the small-world property
of a network would enhance pair correlations between distant objects in
comparison to lattices. However, it is not the case. We demonstrate that
correlations between $\ell $-th nearest neighbors, on average, decay with $%
\ell $ as $1/(\ell z_{\ell })$ or faster. In networks, where the mean intervertex
separation $\overline{\ell }(N)\sim \ln N$, this corresponds to the
exponential decay of correlations with $\ell $ (even at the critical point).
However, in networks with a divergent mean number of the second nearest
neighbors [where $\overline{\ell }(N)$ grows slower than $\ln N$], we
observe a dramatic weakening of pair correlations between the second and
more distant nearest neighbors. In these networks, only the nearest
neighbors are strongly correlated, while the correlations between more
distant vertices are suppressed and approach zero in the infinite network
limit.

In Section \ref{ph} we consider pair correlations in an interacting system
defined on the top of a complex network in the framework of a
phenomenological approach. This approach allows us to understand general
properties of pair correlations irrespective of details of an interacting
system and the nature of interactions. In Section \ref{ising} we support
these results by calculations of the static pair correlation function of the
Ising model on a complex network (more precisely, the configuration model of
a network \cite{bbk72}).

\section{Phenomenological approach}

\label{ph}

Let a quantity $X_{i}(t)$ describe a dynamic process on a network, where
indices $i$ label vertices, and $t$ is time. $x_{i}(t)\equiv
X_{i}(t)-\left\langle X_{i}\right\rangle $ describes fluctuations around the
average value $\left\langle X_{i}\right\rangle $. If the system under
consideration is in an equilibrium state, then pair correlations between two
arbitrary vertices $i$ and $j$ may be characterized by the following
correlation function:
\begin{eqnarray}
G_{ij}(t_{1},t_{2})
&=&t_{0}^{-1}\int\limits_{0}^{t_{0}}x_{i}(t+t_{1})x_{j}(t+t_{2})\,dt  \notag
\\[5pt]
&\equiv &\left\langle x_{i}(t_{1})x_{j}(t_{2})\right\rangle .
\label{c-function}
\end{eqnarray}%
In the present section the brackets $\langle ...\rangle $ mean the average
over the observation time $t_{0}$ which must be much larger than the maximum
relaxation time in the system under consideration. In the equilibrium state
we have $G_{ij}(t_{1},t_{2})=G_{ij}(t_{1}-t_{2})$ in the limit $%
t_{0}\rightarrow \infty $. Assuming the Hamiltonian dynamics, one can
introduce a generalized field $H_{i}(t)$ conjugated with $X_{i}(t)$. The
function%
\begin{eqnarray}
\chi _{ij}(t_{1}-t_{2}) &=&t_{0}^{-1}\int\limits_{0}^{t_{0}}\partial
x_{i}(t+t_{1})/\partial H_{j}(t+t_{2})\,dt  \notag \\
&\equiv &\left\langle \partial x_{i}(t_{1})/\partial
H_{j}(t_{2})\right\rangle  \label{susc}
\end{eqnarray}%
is a generalized non-local susceptibility which characterizes the averaged
response of $x_{i}(t_{1})$ at time $t_{1}$ on a field applied at a vertex $j$
at time $t_{2}$. There is a simple relationship between $G_{ij}(t)$ and $%
\chi _{ij}(t)$:%
\begin{equation}
G_{ij}(t)\varpropto \chi _{ij}(t),  \label{c-s}
\end{equation}%
where coefficient of the proportionality is unimportant for our purpose.

In general case, the total susceptibility $\chi (t)$ of a system (per
vertex) is finite in the limit $N\rightarrow \infty $ where $N$ is the
number of vertices in a network:%
\begin{equation}
\chi (t_{1}-t_{2})=\frac{1}{N}\sum\limits_{i=1}^{N}\left. \left\langle \frac{%
\partial x_{i}(t_{1})}{\partial H_{j}(t_{2})}\right\rangle \right\vert
_{H_{1}=H_{2}=...=H}=O(1).  \label{t-s}
\end{equation}
Here, a statement $A=O(1)$ means that a quantity $A$ is finite in the limit $%
N\rightarrow \infty $. The susceptibility $\chi (t)$ may diverge only at a
critical point. We rewrite Eq. (\ref{t-s}) as follows: 
\begin{equation}
\chi (t)=\frac{1}{N}\sum\limits_{i}\biggl( \chi
_{ii}(t)+\!\!\sum\limits_{j:\ell _{ij}=1}\!\!\!\chi
_{ij}(t)+\!\!\sum\limits_{j:\ell _{ij}=2}\!\!\!\chi _{ij}(t)+...\biggr) ,
\label{t-s2}
\end{equation}
where the first sum in the parentheses is a sum over the nearest neighbors
of a\emph{\ }vertex $i$ being at the distance $\ell _{ij}=1$. The second sum
is over the\emph{\ }second nearest neighbors, $\ell _{ij}=2$, and so on. One
can introduce the average value $\chi (t,\ell )$, that is the average value
of the susceptibility $\chi _{ij}(t)$ at $\ell _{ij}=\ell $ over all
possible network configurations. In the large network limit, this quantity
coincides with the following expression calculated for a single network: 
\begin{equation}
\chi (t,\ell )=\frac{\sum_{i,j:\ell _{ij}=\ell }\chi _{ij}(t)}{%
\sum_{i,j:\ell _{ij}=\ell }1}=\frac{\sum_{i,j:\ell _{ij}=\ell }\chi _{ij}(t)%
}{Nz_{\ell }}\,,  \label{t-s2a}
\end{equation}%
where the sums are over all pairs of vertices at the intervertex distance $%
\ell $, $z_{\ell }=N^{-1}\sum_{i,j:\ell _{ij}=\ell }1$ is the mean number of 
$\ell $-th nearest neighbors. Consequently, 
\begin{equation}
\chi (t)=\sum_{\ell }z_{\ell }\,\chi (t,\ell )\,,  \label{t-s2b}
\end{equation}%
where $z_{0}=1$ and $\chi (t,0)$ is the average local susceptibility. Let us
first assume that all $\chi (t,\ell )\geq 0$. In order to satisfy the
condition $\chi (t)=O(1)$, the terms of the sum in Eq.~(\ref{t-s2b}) must
decay with $\ell $ faster than $1/\ell $. This leads to the following
restriction:\emph{\ }%
\begin{equation}
\chi (t,\ell )<O\!\left( \frac{1}{\ell \,z_{\ell }}\right) \,.  \label{t-s2c}
\end{equation}%
It is important that here $z_{\ell }$ is a function of the network size $N$,
so formula (\ref{t-s2c}) shows how the nonlocal susceptibility (and spacial
correlations) vary with $N$.

If the nonlocal susceptibilities $\chi _{ij}(t)$ are of varying sign,
formula (\ref{t-s2c}) is not applicable. In this case, for the
root-mean-square with averaging over all pairs of vertices $i,j$ with a
given $\ell _{ij}=\ell $, we arrive at 
\begin{equation}
\left[ \overline{\chi _{ij}^{2}(t,\ell _{ij}=\ell )}\right] ^{1/2}<O(z_{\ell
}^{-1/2})\,.  \label{t-s2d}
\end{equation}

In accordance to Eq. (\ref{c-s}), the $\ell $ dependence of the average
correlation function is the same as for the average susceptibility, i.e., it
is described by formulas (\ref{t-s2c}) or (\ref{t-s2d}).

In sparse networks, the mean number of the nearest neighbors (the mean
degree) $z_{1}\equiv \overline{k}=\sum kP(k)$ is finite. In general, in
networks, where the mean intervertex distance $\overline{\ell }(N)\sim N$,
the mean numbers of $\ell $-th nearest neighbors grow exponentially with $%
\ell $. In particular, in uncorrelated networks (without degree--degree
correlations between nearest neighbor vertices), $z_{\ell
}=z_{1}(z_{2}/z_{1})^{\ell -1}$, where $z_{2}=\sum k(k-1)P(k)$ \cite{nsw01}.
In this case, even at the critical point of a cooperative model, the
nonlocal susceptibility and pair correlations decrease exponentially rapidly
with $\ell $.

The effect is even more dramatic, if the degree distribution is fat-tailed,
and its second moment diverges as $N\rightarrow \infty $. In scale-free
networks, this takes place if the $\gamma $ exponent of the degree
distribution $P(k)\sim k^{-\gamma }$ is equal or below $3$. In this case,
already the mean number $z_{2}$ of the second nearest neighbors of a vertex
diverges as $N\rightarrow \infty $. Consequently, according to Eq.~(\ref%
{t-s2c}) or Eq.~(\ref{t-s2d}), in the limit of a large network, pair
correlations between the second (and further) nearest neighbors vanish, and
only pair correlations between the nearest neighbors are observable.

\section{Correlations in the Ising model}

\label{ising}

The simplicity of the Ising model and the fact that many microscopic models
may be mapped to this model make the Ising model to be very attractive.
Recent investigations have revealed that the critical behavior of the Ising
and Potts models on complex networks strongly differs from the standard mean-field
behavior on a regular lattice \cite{ahs01,dgm02,lvvz02,b02,ceah02,lgkk04,em04}. In the present
section we analyze pair correlations in the equilibrium Ising model on an
uncorrelated random complex network.

\subsection{The model}

\label{model}

We consider the ferromagnetic Ising model: 
\begin{equation}
\mathcal{H}=-J\sum_{\langle ij\rangle }S_{i}S_{j}-\sum_{i}H_{i}S_{i}\,,
\label{H}
\end{equation}%
%
%
%
%
%
%
%
where $S_{i}=\pm 1$, and $H_{i}$ is a local magnetic field at a vertex $i$.
The sum is over all nearest neighboring vertices. The static
pair-correlation function 
\begin{equation}
G_{ij}\equiv \left\langle S_{i}S_{j}\right\rangle -\left\langle
S_{i}\right\rangle \left\langle S_{j}\right\rangle  \label{1c}
\end{equation}%
is related to the non-local magnetic susceptibility $\chi _{ij}$:%
\begin{equation}
\chi _{ij}=\partial M_{i}/\partial h_{j}=\beta G_{ij},  \label{xi}
\end{equation}%
where $\beta =1/T$ and $T$ is temperature, $M_{i}\equiv \left\langle
S_{i}\right\rangle $. In this section $\left\langle ...\right\rangle $ means
the statistical average with the Hamiltonian $\mathcal{H}$.

As a substrate, we use the standard model of an uncorrelated random
network---the configuration model \cite{bbk72}. This is the maximally random
graph with a given degree distribution or, as it is called in graph theory,
a labelled random graph with a given degree sequence. It is important that
the uncorrelated networks have a locally tree-like structure. That is, in
the large network limit, there are no loops in a finite neighborhood of a
vertex. These networks may be considered as the random Bethe lattices,
which, by definition, have no boundary. One should note that in contrast,
the Cayley tree contains boundary vertices which are dead ends \cite{bbook82}%
.\emph{\ }

\subsection{\!\!How to solve the Ising model on a tree-like graph}

\label{model1}

Let us outline an effective approach to solution of cooperative models on
tree-like networks, which we use for obtaining $\chi _{ij}$. This method was
implemented for the Ising and Potts models on regular \cite{bbook82} and random \cite{dgm02,dgm04} Bethe
lattices. Consider an arbitrary tree-like graph.
Consider a spin $S_{i}$ on a vertex $i$ of this graph with $k_{i}$ adjacent
spins $S_{j}$, $j=1,2,\ldots ,k_{i}$. As any vertex in this graph, vertex $i$
may be treated as a root of a tree. In turn, vertices $j$ may be treated as
roots of subtrees growing from the vertex $i$. In order to characterize the
subtree with the root spin $S_{j}$ we introduce a quantity 
\begin{eqnarray}
g_{ij}(S_{i}) &=&\sum_{\{S_{l}\}=\pm 1}\exp \biggl[\beta J\sum_{\left\langle
nm\right\rangle }S_{n}S_{m}  \notag \\
&&+\beta JS_{i}S_{j}+\beta \sum_{n}H_{n}S_{n}\biggr].  \label{gg1}
\end{eqnarray}%
Here the indices $n$ and $m$ run only over spins that belong to the
sub-tree, including $S_{j}$. The statistical sum in Eq. (\ref{gg1}) is taken
only over the spins on the subtree.

We introduce a quantity 
\begin{equation}
x_{ij}\equiv g_{ij}(-1)/g_{ij}(+1)\,.  \label{xij}
\end{equation}%
In such a way, for each vertex $i$ we can introduce $k_{i}$ parameters $%
x_{ij}$, $j=1,2,...k_{i}$. It should be noted that $x_{ij}\neq x_{ji}$
because $x_{ij}$ and $x_{ji}$ characterize different sub-trees. So, each
edge is characterized by a pair: $x_{ij}$ and $x_{ji}$. In sum, the Ising
model on an arbitrary tree-like graph is described by $2L=\sum%
\nolimits_{i}k_{i}$ \ parameters $x_{ij}$. Here, $L$ is the total number of
edges of the graph.

Using Eqs. (\ref{gg1}) and (\ref{xij}), the parameter $x_{ij}$ may be
related to parameters $x_{jl}$ which characterize the edges (sub-trees)
outgoing from the vertex $j$ as follows \cite{bbook82}: 
\begin{equation}
x_{ij}=y\Biggl(\!H_{j,}\prod_{l=1}^{k_{j}-1}x_{jl}\Biggr)\,,  \label{recurr}
\end{equation}%
where $l$ is the index of the first nearest neighbors of the vertex $j$. The
vertex $i$ is excluded from the product over $l$. In other words, a vertex $%
l $ is a second neighbor of the vertex $i$ and the first one of the vertex $%
j $. The function $y(H,x)$ in Eq.~(\ref{recurr}) depends on a cooperative
model. For the Ising model,%
\begin{equation}
y(H,x)=\frac{e^{(-J+H)\beta }+e^{(J-H)\beta }x}{e^{(J+H)\beta
}+e^{(-J-H)\beta }x}\,.  \label{y(x)}
\end{equation}%
$2L$ independent parameters $x_{ij}$ should be found by solution of the set
of $2L$ equations (\ref{recurr}) at a given temperature $T$ and local
magnetic fields $H_{i}$. For an arbitrary tree-like graph, these equations
may be solved numerically, e.g., by use of the population dynamic algorithm,
see\emph{\ }Ref. \cite{em04} where this method has been applied to the Potts
model on a tree-like graph.

Observable thermodynamic quantities of the Ising model may be written as
functions of the parameters $x_{ij}$. For example, a magnetic moment $M_{i}$
is given by the following equation:%
\begin{equation}
M_{i}=\Biggl(e^{2\beta H_{i}}-\prod_{j=1}^{k_{i}}x_{ij}\Biggr)\!\!\Biggm/\!\!\Biggl(e^{2\beta
H_{i}}+\prod_{j=1}^{k_{i}}x_{ij}\Biggr)\,.  \label{M}
\end{equation}%
Finally, for a random network, the resulting observables should be averaged
over various configurations with appropriate statistical weights.

\subsection{Derivation of the pair correlation function}

\label{model2}

Let us find a non-local susceptibility $\chi _{ij}$ for two arbitrary
vertices $i$ and $j$ at the distance $l_{ij}=\ell $ from each other. On a
tree-like graph there is the only shortest way connecting these vertices. It
starts from $i$, then goes through vertices $i_{1}$, $i_{2},\ldots ,i_{\ell
-1}$ and ends at $j$. Using Eqs. (\ref{M}) and (\ref{recurr}) we get 
\begin{equation}
\chi _{ij}=\frac{\partial M_{i}}{\partial x_{ii_{1}}}\frac{\partial
x_{ii_{1}}}{\partial x_{i_{1}i_{2}}}\frac{\partial x_{i_{1}i_{2}}}{\partial
x_{i_{2}i_{3}}}...\frac{\partial x_{i_{l-2}i_{\ell -1}}}{\partial x_{i_{\ell
-1}j}}\frac{\partial x_{i_{\ell -1}j}}{\partial H_{j}}.  \label{sij}
\end{equation}%
If we know $x_{ij}$ we can get $\chi _{ij}$.

First, for the purpose of comparison, we find a non-local susceptibility $%
\chi _{ij}$ of a regular Bethe lattice with a coordination number $k$.

In a uniform magnetic field $H_{1}=H_{2}=....=H$ all vertices and all edges
in a regular Bethe lattice are equivalent. Therefore, the parameters $x_{ij}$
are equal, i.e. $x_{ij}=x$, and Eq. (\ref{recurr}) is reduced to 
\begin{equation}
x=y(H,x^{k-1})\,.  \label{x-bethe}
\end{equation}%
This equation determines $x$ as a function of $T$ and $H$. From Eqs. (\ref%
{sij}) and (\ref{x-bethe}) we get%
\begin{equation}
\chi (\ell )=\frac{1}{k}\frac{\partial M}{\partial x}\left[ \frac{1}{(k-1)}%
\frac{\partial y(H,x^{k-1})}{\partial x}\right] ^{\ell -1}\frac{\partial
y(H,x^{k-1})}{\partial H}\,,  \label{bethe}
\end{equation}%
where in accordance with Eq. (\ref{M}) we have $M=(e^{2\beta
H}-x^{k})/(e^{2\beta H}+x^{k})$.\ At zero magnetic field $H=0$, Eq. (\ref%
{bethe}) takes a form:%
\begin{equation}
\chi (\ell )=\frac{4\beta x^{k}}{(1+x^{k})^{2}}\left[ \frac{2x^{k-2}\sinh
(2J\beta )}{(e^{J\beta }+e^{-J\beta }x^{k-1})^{2}}\right] ^{\ell }\,.
\label{xg}
\end{equation}%
This equation is valid at all temperatures.

In the paramagnetic state, i.e. at temperatures $T$ above the critical
temperature $T_{c}=2J/\ln [k/(k-2)]$ of the ferromagnetic phase transition,
the self-consistent equation (\ref{x-bethe}) has the only solution $x=1$,
and we get%
\begin{equation}
\chi (\ell )=\beta G(l)=\beta \tanh ^{\ell }(J\beta )  \label{xb}
\end{equation}%
in agreement with the result obtained in Ref. \cite{f75} for a Cayley tree
in the framework of another method. Furthermore, using the fact that $%
x_{ij}=1$ at $T\geq T_{c}$, one can show that in the paramagnetic phase, $%
\chi (\ell )$ of the Ising model on an arbitrary tree-like graph is the same
as that for an arbitrary Cayley tree, e.g., for a spin chain (see below).
Recall that there is no phase transition in spin models on a Cayley tree due
to the presence of boundary spins. The number of these spins is of the order
of the total number of spins on a given Cayley tree.

In the ferromagnetic phase, i.e. at $T<T_{c}$, or at $H\neq 0$ the parameter 
$x$ is smaller than 1. 
Note that in accordance with Eqs. (\ref{xg}) and (\ref{xb}) the non-local
susceptibility $\chi (\ell )$ of a regular Bethe lattice is finite at all
temperatures and magnetic fields.

Now let us consider an uncorrelated random network. We calculate a value of $%
\chi _{ij}(\ell _{ij}=\ell )$ averaged over the ensemble of uncorrelated
random graphs with a given degree distribution function $P(k)$: 
\begin{eqnarray}
&&\chi (\ell )\equiv \overline{\chi _{ij}(\ell _{ij}=\ell )}%
=\sum\limits_{k_{i}}\sum\limits_{k_{1}}...\sum\limits_{k_{l-1}}\sum%
\limits_{k_{j}}\chi _{ij}  \notag \\[5pt]
&&\!\!\!\!\!\!\!\!\!\!\!\!\!\times \left[ \frac{P(k_{j})k_{j}}{z_{1}}\left\{
\prod\limits_{n=1}^{\ell -1}\frac{P(k_{n})k_{n}(k_{n}-1)}{z_{2}}\right\} 
\frac{P(k_{i})k_{i}}{z_{1}}\right] \,.  \label{ax}
\end{eqnarray}%
Here, the quantity in the square brackets is the probability that a vertex $%
i $ with degree $k_{i}$ is connected with a vertex $j$ having degree $k_{j}$
by a path that goes through vertices with degrees $k_{1}$, $k_{2}$, ...$%
k_{\ell -1}$.

In the paramagnetic phase at $T\geq T_{c}=2J/\ln [\langle k^{2}\rangle
/(\langle k^{2}\rangle -2\langle k\rangle )]$ and $H=0$ we have $x_{ij}=1$,
and Eq. (\ref{ax}) leads to the Eq. (\ref{xb}). Therefore, in zero magnetic
field both on regular and random Bethe lattices the non-local susceptibility
and $\chi _{ij}$ have the same temperature dependence determined by Eq. (\ref%
{xb}).

At a magnetic field and in the ferromagnetic phase an approximate expression
for the function $\chi (\ell )$ may be obtained in the framework of the
approach proposed in \cite{dgm02}. We introduce $x_{ij}=\exp (-h_{ij})$.
Here, the quantities $h_{ij}$ are positive and independent random
parameters. We use the following ansatz,

\begin{equation}
\sum_{j=1}^{k}h_{ij}\approx kh+\mathcal{O}(k^{1/2})\,,  \label{approx}
\end{equation}%
where $h\equiv \overline{h_{ij}}$ is the average value of the parameter.
Applying the ansatz (\ref{approx}) to Eq. (\ref{recurr}) yields

\begin{equation}
h=-\frac{1}{\overline{k}}\sum_{k}P(k)k\ln y(H,e^{-(k-1)h})  \label{h-av}
\end{equation}%
which determines $h$ as a function of $T$ and $H$. In fact, the parameter $h$
plays the role of the order parameter. At $H=0$, we get $h=0$ above $T_{c}$
and non-zero below $T_{c}$.

With this ansatz, 
\begin{equation}
\chi (\ell )=z_{1}^{2}z_{2}^{-(\ell -1)}AB^{\ell -1}C\,,  \label{2c}
\end{equation}%
where 
\begin{eqnarray}
&&A=-\sum_{q}qP(q)\frac{2e^{-qh}}{(e^{H\beta }+e^{-H\beta -qh})^{2}}\,,
\label{3c} \\[5pt]
&&\!\!\!\!\!\!\!\!\!\!\!\!\!\!\!\!\!B=\sum_{q}q(q\!-\!1)P(q)\frac{%
2e^{-(q-2)h}\sinh (2J\beta )}{(e^{(J+H)\beta }\!+\!e^{-(J+H)\beta
-(q-1)h})^{2}},  \label{5c} \\[5pt]
&&\!\!\!\!\!\!\!\!\!\!\!\!\!C=-\sum_{q}qP(q)\frac{4e^{-(q-1)h}\sinh (2J\beta
)}{(e^{(J+H)\beta }+e^{-(J+H)\beta -(q-1)h})^{2}}\,.  \label{4c}
\end{eqnarray}%
%
%
%
%
%
%
%
The quantities $A,B$ and $C$ are finite (independent of $N$) for an
arbitrary degree distribution $P(k)$. The finiteness of the sums in Eqs.~(%
\ref{3c})--(\ref{4c}) is ensured by the exponential multiplier $e^{-qh}$.

Result (\ref{2c}) is in agreement with the conclusions of the preceding
section [compare relations (\ref{2c}) and (\ref{t-s2c})]. Note that the
conclusion that $\chi (\ell \geq 2)\propto G(\ell \geq 2)\rightarrow 0$ as $%
N\rightarrow \infty $ if the second moment of the degree distribution
diverges,\emph{\ }does not depend on the fact that we used the approximation
(\ref{approx}). Indeed, substituting Eq. (\ref{sij}) into Eq. (\ref{ax}),
one can prove that all the sums over degrees $k_{n}$ converge both in the
ordered state and at $H\neq 0$. Therefore, we get $\chi (\ell )\propto
z_{2}^{-(\ell -1)}$ in agreement with Eq. (\ref{2c}).

\section{Discussion and conclusions}

\label{concl}

In the framework of a phenomenological approach and by using the Ising model
we have studied time and static pair correlations $\left\langle
x_{i}(t_{1})x_{j}(t_{2})\right\rangle $ between interacting objects defined
on a random uncorrelated complex network. We have demonstrated that if a
complex network has a divergent number of the second neighbors, then the
pair correlations between second and more distant neighbors are strongly
suppressed in large networks. In fact, the correlations tend to zero in the
limit $N\rightarrow \infty $. One should note that a particular form of the
characteristic $z_{2}(N)$ strongly depends on a network model. It is
important that the size dependence of the pair correlation function is
determined by the factor $1/z_{2}^{\ell -1}(N)$.

The effect of weakening of the pair correlations is well known to people
leaving in large towns and forming a large network of acquaintances. Close
relationships are only being kept between close relatives inside family,
close friends or between people working in the same office. These people are
our nearest neighbors. Our \textquotedblleft second\textquotedblright\ and
more distant nearest neighbors---people living on the next floor or working
in a neighboring office ---usually weakly influence our personal and social
life. This\emph{\ }is contrary to a small village where strong correlations
are present between almost all inhabitants.

Our analysis may be generalized to correlated networks. Many natural
networks exhibit correlations between degrees of adjacent vertices, see, for
example, Ref. \cite{corr}. Note that in large uncorrelated networks, the
divergence of the mean number $z_{\ell }$ of the $\ell $-th nearest neighbor
can occur only simultaneously at all $\ell \geq 2$. In contrast, in large
correlated networks, it is possible in principle that, say, $z_{2},z_{3}$ is
finite and only $z_{\ell \geq 4}$ diverges. In this example, pair
correlations are observable between the first, second and third nearest
neighbors and vanish starting from the fourth nearest neighbors.\emph{\ }

We restricted ourselves to pair correlations. One should emphasize that pair
correlations in cooperative systems on networks, unlike lattices, never show
critical behavior. Even at the critical point of an interacting system on a
network, we observe an exponentially rapid decay of pair correlations. In
contrast, volume correlations in interacting systems on networks demonstrate
a critical feature. Let us explain this point. The distribution of the full
response of a system to a small local field is $P(\varepsilon
)=\sum_{i}\delta (\sum_{j}\chi _{ij}-\varepsilon )$, where $\delta
(\varepsilon )$ is the delta-function. This distribution is a rapidly
decreasing function both below and above a phase transition on a network. On
the other hand, at the critical point, $P(\varepsilon )$ is power-law, that
is, \textquotedblleft critical\textquotedblright\ \cite{nsw01,bcd05}.

One can conclude that a network structure of an interacting system strongly
influences the dependence of pair correlations on the distance between
interacting objects. So, investigations of the correlations may be a useful
method to understand the network topology. If an interacting system is
defined on a network with a divergent number $z_{n}$ of $n$-th nearest
neighbors of a vertex, then dynamic and static pair correlation are strongly
suppressed at distances $n$ and greater.

This work was partially supported by projects POCTI: FAT/46241/2002,
MAT/46176/2002, FIS/61665/2004, and BIA-BCM/62662/2004. S.N.D. and J.F.F.M.
were also supported by project DYSONET-NEST/012911. Authors thank
A.N.~Samukhin for useful discussions.

\end{document}